\documentclass[twocolumn,
prl
,superscriptaddress
]{revtex4-1}

\usepackage{graphicx}
\usepackage{amssymb}
\usepackage{amsmath}
\usepackage[usenames,dvipsnames]{color}
\usepackage{comment}

\DeclareMathOperator{\tr}{tr}

\newcommand{\vect}[1]{{\mathbf #1}}
\newcommand{\Frac}[2]{\displaystyle\frac{#1}{#2}}
\newcommand{\MF}{\text{MF}}

\begin{document}


\title{Collective pairing of resonantly coupled microcavity
  polaritons}

\author{F. M. Marchetti} 
\email{francesca.marchetti@uam.es} 
\affiliation{Departamento de F\'isica Te\'orica de la Materia
  Condensada \& Condensed Matter Physics Center (IFIMAC), Universidad
  Aut\'onoma de Madrid, Madrid 28049, Spain}

\author{Jonathan Keeling} 
\email{jmjk@st-andrews.ac.uk} 
\affiliation{SUPA, School of Physics
  and Astronomy, University of St Andrews, St Andrews KY16 9SS UK}

\date{\today}

\begin{abstract}
  We consider the possible phases of microcavity polaritons tuned near
  a bipolariton Feshbach resonance.  We show that, as well as the
  regular polariton superfluid phase, a ``molecular'' superfluid
  exists, with (quasi-)long-range order only for pairs of polaritons.
  We describe the experimental signatures of this state. Using
  variational approaches we find the phase diagram (critical
  temperature, density and exciton-photon detuning).  Unlike ultracold
  atoms, the molecular superfluid is not inherently unstable, and our
  phase diagram suggests it is attainable in current experiments.
\end{abstract}


\maketitle

%
The wealth of physics explored with ultracold atomic
gases~\cite{Bloch2008a} relies on the ability to tune parameters such
as the interaction strength. A crucial tool to achieve this is the
Feshbach resonance mechanism~\cite{Chin2010}: By using a magnetic
field one may vary the detuning $\nu$ between two channels (hyperfine
states) of the atoms; a closed channel (bound molecule) and open
(scattering) channel.  When $\nu$ is large and positive, the closed
channel is far above open channel atoms, the formation of molecules is
energetically suppressed and atoms scatter with a weakly attractive
effective interaction.  When $\nu$ is large and negative, atoms are
paired into molecules and the effective interaction is weakly
repulsive.  Near resonance ($\nu \simeq 0$), the interaction is very
large. This enables tunable pairing and regimes of strong
correlations, allowing many interesting possibilities both at few- and
many-body level.

Ultracold atom experiments are however intrinsically metastable, not
true minima of the free energy. Three-body collisions must be avoided
to prevent relaxation to, e.g., a solid phase. Also, Feshbach
molecular states are highly rovibrationally excited states and can
relax to lower states.  Exploring regimes of strong interactions while
suppressing such relaxation processes is inherently challenging.
For fermionic atoms, Pauli exclusion suppresses scattering rates, so
strongly interacting regimes can be accessed.
In this case there is a smooth crossover between a
Bardeen-Cooper-Schrieffer (BCS) condensate of weakly attractive
fermions and a Bose-Einstein condensate (BEC) of repulsive
molecules~\cite{W.Ketterle,Leggett2012}.

The situation for bosons near a Feshbach resonance differs
substantially: Unlike fermions, both a condensate of molecules and a
condensate of unpaired bosons can exist.  As discussed below this
means that a ``molecular'' superfluid phase with no off-diagonal
long-range order (ODLRO) for atoms can arise, with a further symmetry
breaking phase transition between ``atomic'' and ``molecular''
superfluids.  Bosonic mixtures have attracted considerable theoretical
interest~\cite{Timmermans:Feshbach,Mueller2000,Jeon2002,kuklov04_1,kuklov04_2,Radzihovsky2004,Romans2004a,Radzihovsky2008a,Basu:Stability,Zhou2008a,Koetsier2009,Bhaseen:Polaritons,Radzihovsky2009,Hohenadler:QPT,Ejima2011,Bhaseen2012a,Zhou2013}.
However, experiments have been limited by stability issues: The
molecular superfluid phase generally requires high
densities~\cite{Basu:Stability,Koetsier2009}, where three body losses
are significant~\cite{Rem2013}.  There have been suggestions to use
optical lattices~\cite{kuklov04_1} to reach the strongly interacting
regime while avoiding high densities, but to date, molecular pairing
phases of bosonic atoms remain elusive.

Microcavity polaritons~\cite{kavokin_laussy,Carusotto2013a}, the
quasiparticles resulting from strong coupling between cavity photons
and quantum well excitons, do not suffer the metastability problems of
cold atoms and so present a more promising venue to study bosonic
pairing phases. Microcavity polaritons have been observed to form a
BEC~\cite{Kasprzak2006,snoke07science}, and, as recently
discussed~\cite{Wouters2007,Carusotto2010a}, interactions between
polaritons with opposite polarizations can support either a bound
state (bipolariton) or a scattering resonance depending on the
exciton-photon detuning. Several signatures of this physics have been
seen~\cite{Saba2000,Borri2003,Wen2013}, including recent direct
observations of the scattering resonance vs
detuning~\cite{takemura2014,Takemura2013}.  Polaritons are expected
not to suffer from the issues of three body inelastic
losses~\cite{Basu:Stability,Koetsier2009,Rem2013} which plague cold
bosonic atoms:
Unlike ultracold atoms, a polariton BEC is not a metastable state,
rather it is the minimum free energy state as long as the polariton
population is conserved.  In addition, there is no deeply bound
molecular state below the bipolariton.  
Polaritons do have a finite lifetime, however recent experiments have
demonstrated a 5--10 fold increase in lifetime~\cite{Wen13}, leading
to a system very close to thermal equilibrium.  A full non-equilibrium
treatment of spin dynamics and relaxation is beyond the scope of this
work. However, unlike cold atoms, such losses are not intrinsically
linked to the resonance physics, and so they can be addressed and
improved independently.  Thus, polaritons may form a more promising
venue to study pairing phases.

In this Letter we explore the phase diagram of collective paired
phases arising from a bipolariton resonance in microcavities.  By
using the cavity-exciton detuning to tune the
interactions~\cite{Wouters2007,Carusotto2010a}, we reveal a phase
transition between atomic (i.e. polariton) and molecular
(i.e. bipolariton) BEC phases. We show that temperatures and detunings
required for typical materials such as GaAs are attainable and we
discuss experimental signatures for detecting such phases.

\begin{figure}
\centering
\includegraphics[width=3.2in]{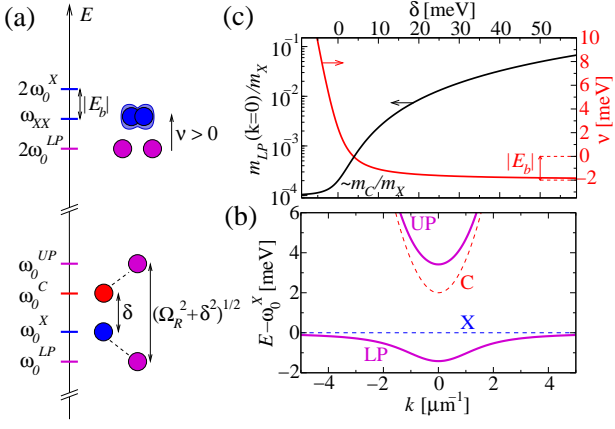}
\caption{ (Color online) (a) Cartoon of one- and two-particle energy
  scales, showing how the cavity(C)--exciton(X) detuning $\delta$
  determines the inter-channel detuning $\nu$. (b) Lower (LP) and
  upper (UP) polariton dispersions (solid [magenta] lines), along with
  the C (X) dispersion (dashed [red (blue)] lines) for
  $\delta=2$~meV. (c) Dependence of the rescaled polariton mass at
  zero momentum $m_{LP} (\vect{k}=0)/m_X$ (left [black]) and of the
  inter-channel detuning $\nu$ (right [red]) on $\delta$. Plots are
  for GaAs parameters ($\Omega_R=4.4$~meV,
  $|E_b|=2$~meV~\cite{Wouters2007}, photon mass $m_C=10^{-4}m_X$,
  exciton mass $m_X=0.4 m_e$).}
\label{fig:dispe}
\end{figure}
%
\paragraph{Model ---}
In the model we consider, bipolaritons play the r\^ole of closed
channel molecules, while the two (polarized) LP modes are the open
channel modes.
The inter-channel detuning is (see Fig.~\ref{fig:dispe}), with $E_b$
the biexciton binding energy and $2 \omega_{\vect{k}}^{LP} =
\omega_{\vect{k}}^{C} + \omega_{\vect{k}}^{X} -
\sqrt{(\omega_{\vect{k}}^{C} - \omega_{\vect{k}}^{X})^2 +
  \Omega_R^2}$, where both the exciton and cavity photon dispersions
are quadratic, $\omega^{C,X}_{\vect{k}} = \omega^{C,X}_{\vect{0}} +
\frac{k^2}{2m_{C,X}}$.
In fact, as discussed later and in~\cite{SM}, the bipolariton bound
state is almost identical to a biexciton. However, it possesses a
non-zero photon component and for this reason we refer to it as a
bipolariton. Even a small photon content has important consequences in
breaking the degeneracy between dark and bright excitonic
states~\cite{SM}.
The inter-channel detuning $\nu$ can be controlled by varying the
cavity-exciton detuning $\delta = \omega_0^C - \omega_0^X$ (see
Fig.~\ref{fig:dispe}). A resonant enhancement of the interactions
occurs near the bare resonance $\nu \simeq 0$.

We derive the many-body properties of resonantly coupled polaritons by
considering a two-channel model, which includes both LPs in the right-
and left-circular polarization basis,
$\hat{\psi}_{\uparrow,\downarrow}$ and bipolariton $\hat{\psi}_m$
fields.
One could, alternatively, work with a single-channel model with no
explicit bipolariton field, at the expense of needing a finite range
attractive potential $U^{\uparrow \downarrow} (\vect{r})$ supporting a
resonant bound state~\cite{Gurarie2007}. Such an approach
unnecessarily complicates the derivation of many-body physics, while a
single-channel model with a contact potential cannot describe deeply
bound bipolariton states.

The effective two-channel polariton model has been derived starting
from a model describing coupled exciton, biexction and photon fields
and rotating to the LP $\hat{\psi}_{\uparrow,\downarrow}$ and
bipolariton $\hat{\psi}_m$ fields (see the scheme in
Fig.~\ref{fig:dispe}~(a)). This yields the following
grand-canonical Hamiltonian written in momentum space,
$\hat{\psi}_{\sigma} (\vect{r}) = \sum_{\vect{k}} e^{i \vect{k} \cdot
  \vect{r}} \hat{a}_{\vect{k} \sigma}/\sqrt{\mathcal{A}} $ (where
$\vect{k}$ is the in-plane [2D] momentum, $\mathcal{A}$ the system
area, and $\hbar=1$ throughout):
\begin{multline}
  \hat{H} = \sum_{\sigma= \uparrow, \downarrow, m} \sum_{\vect{k}}
  \hat{a}_{\vect{k} \sigma}^\dag \left(\epsilon_{\vect{k} \sigma} -
  \mu_{\sigma}\right) \hat{a}_{\vect{k} \sigma}^{} \\ +
  \Frac{1}{\mathcal{A}} \sum_{\sigma \sigma'} \sum_{\vect{k},
    \vect{k}', \vect{q}} \Frac{U_{\vect{k} \vect{k}' \vect{q}}^{\sigma
      \sigma'}}{2} \hat{a}_{\vect{k} \sigma}^\dag \hat{a}_{\vect{k}'
    \sigma'}^\dag \hat{a}_{\vect{k}' - \frac{\vect{q}}{2} \sigma'}^{}
  \hat{a}_{\vect{k}+ \frac{\vect{q}}{2} \sigma}^{} \\ +
  \Frac{1}{\sqrt{\mathcal{A}}} \sum_{\vect{k}, \vect{Q}}
  \Frac{g_{\vect{k} \vect{Q}}}{2} \left(\hat{a}_{\vect{k} +
    \frac{\vect{Q}}{2} \uparrow}^\dag \hat{a}_{-\vect{k} +
    \frac{\vect{Q}}{2} \downarrow}^\dag \hat{a}_{\vect{Q} m}^{}
  +\text{h.c.}\right) \; .
\label{eq:model}
\end{multline}
Here $\epsilon_{\vect{k} \uparrow,\downarrow} = \omega_{\vect{k}}^{LP}
- \omega_{0}^{LP}$ is the full LP dispersion (see
Fig.~\ref{fig:dispe}~(b)).  The coupling $g_{\vect{k} \vect{Q}}$
between open and closed channels, and the closed channel (bipolariton)
dispersion can be derived by including the exciton-photon coupling in
the exciton $T$-matrix, following~\cite{Wouters2007} (see~\cite{SM}
for details).  Because $m_C \ll m_X$, coupling to photons only weakly
renormalizes the scattering resonance properties, hence, the
bipolariton and biexciton dispersions almost coincide, so that
$\epsilon_{\vect{k} m} = \frac{k^2}{2 m_m}$ with $m_m \simeq 2
m_X$. In the absence of a magnetic field, the $\uparrow$ and
$\downarrow$ populations are equal and the effective chemical
potentials are $\mu_{\uparrow} = \mu_{\downarrow} = \mu$ and $\mu_m =
2\mu - \nu$, with $\nu$ as discussed above. The case with
$\mu_{\uparrow} \ne \mu_{\downarrow}$ will be the subject of future
study.

To account for the varying excitonic fraction along the LP dispersion,
the interaction terms include the Hopfield coefficient,
$2c_{\vect{k}}^2 = 1 + (\omega_{\vect{k}}^{C} -
\omega_{\vect{k}}^{X})/\sqrt{(\omega_{\vect{k}}^{C} -
  \omega_{\vect{k}}^{X})^2 + \Omega_R^2}$. One finds that $g_{\vect{k}
  \vect{Q}} = c_{\vect{k} + \frac{\vect{Q}}{2}} c_{-\vect{k} +
  \frac{\vect{Q}}{2}} \sqrt{\Delta^X/m_X}$ where $\Delta^X \simeq 4
|E_b|$~\cite{Wouters2007} is the excitonic resonance
width~\cite{Gurarie2007}. The equivalent polaritonic energy scale
$\Delta^{LP}_{\vect{k} \vect{Q}}=m g_{\vect{k} \vect{Q}}^2$ is given
in terms of the LP mass at zero momentum $m = m_{LP} (\vect{k}=0)=
[c_0^2/m_X + (1-c_0^2)/m_C]^{-1}$ (see Fig.~\ref{fig:dispe}~(c)).
Similarly, the polariton interaction strengths also depend on the
Hopfield coefficients~\cite{wouters07}, $U_{\vect{k} \vect{k}'
  \vect{q}}^{\sigma \sigma'} = {\tilde{U}_{\vect{k} \vect{k}'
    \vect{q}}^{\sigma \sigma'}}/{m} = c_{\vect{k}} c_{\vect{k}'}
c_{\vect{k}'-\frac{\vect{q}}{2}} c_{\vect{k} + \frac{\vect{q}}{2}}
{\tilde{U}^{\sigma \sigma'}_X}/{m_X}$, where $\tilde{U}^{\sigma
  \sigma'}$ and $\tilde{U}^{\sigma \sigma'}_X$ are dimensionless
constants. Note that the parameters $U^{\uparrow \uparrow} =
U^{\downarrow \downarrow}$, $2U^{\uparrow \downarrow}$, and $U^{mm}$
are the background (i.e., far from resonance) interaction strengths.
Near resonance, the physical interaction also includes effects of the
hybridization $g_{\vect{k}\vect{Q}}$.  To estimate experimentally
relevant parameters we consider GaAs microcavities (as in
Fig.~\ref{fig:dispe}), for which $\Omega_R=4.4$~meV, $|E_b|=2$~meV
\cite{Wouters2007}, $m_C=10^{-4}m_X$, $m_X=0.4 m_e$, the resonance
$\nu=0$ is at $\delta=3.84$~meV, and $\tilde{U}^{\uparrow \uparrow}_X
= \tilde{U}^{\downarrow \downarrow}_X = 6$ and we take
$\tilde{U}^{\uparrow\downarrow}=0$~\cite{Wouters2007}.  Because $m_C
\ll m_X$, the LP fluid is much more weakly interacting than the
excitonic fluid, i.e., $\tilde{U}^{\sigma \sigma'} \ll
\tilde{U}^{\sigma \sigma'}_X$, except when $\delta \gg \Omega_R$ and
the LP becomes pure exciton.  As widely discussed for
atoms~\cite{Mueller2000,Basu:Stability,Koetsier2009}, repulsive
closed-channel interactions, $\tilde{U}^{mm}_X>0$, are required for
stability.
Biexciton interactions are indeed repulsive, as discussed in the
context of biexciton condensation~see e.g.~\cite{Mysryowicz1995}, and
we take $\tilde{U}^{mm}_X = 4$.

\begin{figure}
\centering
\includegraphics[width=3.2in]{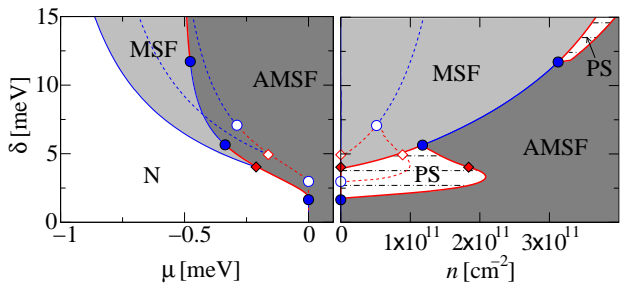}
\caption{(Color online) $T=0$ phase diagram for a GaAs microcavity,
  plotting cavity-exciton detuning $\delta$ vs. chemical potential
  $\mu$ (left) or total density $n$ (right). The normal phase N is
  white, MSF shaded light- and AMSF is dark-grey. First (thick [red])
  and second (thinner [blue]) order lines are separated by tricritical
  points ([blue] circles), while critical end-points are (red)
  diamonds. Horizontal dot-dashed lines connect the same chemical
  potential first-order boundaries in the phase separated (PS)
  region. Dashed lines are the mean-field boundaries neglecting
  quantum fluctuations.}
\label{fig:phase}
\end{figure}
%
\paragraph{Zero temperature ---}
Figure~\ref{fig:phase} shows the $T=0$ phase diagram for the GaAs
parameters specified above. The axes are the detuning $\delta$ and
either the chemical potential $\mu$ or the total density,
$n=\sum_{\vect{k}} \langle \hat{a}_{\vect{k} \uparrow}^{\dag}
\hat{a}_{\vect{k} \uparrow}^{} + \hat{a}_{\vect{k} \downarrow}^{\dag}
\hat{a}_{\vect{k} \downarrow}^{} + 2 \hat{a}_{\vect{k} m}^{\dag}
\hat{a}_{\vect{k} m}^{}\rangle/\mathcal{A}$.  Details of the
calculation appear below. Varying $\delta$ changes both the
inter-channel detuning $\nu$ and the LP dispersion $\omega_k^{LP}$.
Three phases exist~\footnote{For imbalanced populations in presence of
  a Zeeman field, two additional phases are allowed~\cite{Zhou2008a}:
  {$\psi_\uparrow\ne0$, $\psi_\downarrow=0$, $\psi_m = 0$ and
    $\psi_\uparrow = 0$, $\psi_\downarrow\ne0$, $\psi_m = 0$}.}: The
normal phase (N), $\psi_{\uparrow} = \psi_{\downarrow} = \psi_m= 0$
(where $\psi_\sigma = \langle \hat{\psi}_{\sigma}^{} \rangle = \langle
\hat{a}_{\vect{k} \sigma}\rangle \delta_{\vect{k},0}
/\sqrt{\mathcal{A}}$) and two condensed phases arising from the
spontaneous symmetry breaking of the $U(1) \times U(1)$ global
symmetry, $\hat{\psi}_{\uparrow ,\downarrow} \mapsto
e^{i\phi_{\uparrow ,\downarrow}}\hat{\psi}_{\uparrow ,\downarrow}$,
$\hat{\psi}_{m} \mapsto e^{i (\phi_{\uparrow} +
  \phi_{\downarrow})}\hat{\psi}_{m}$.  Condensation of atoms,
$\psi_{\uparrow} = \psi_{\downarrow} \ne 0$, guarantees that of
molecules, $\psi_m\ne 0$ (but not vice-versa). We denote this phase as
an atomic and molecular superfluid phase (AMSF): Here, the $U(1)
\times U(1)$ symmetry is completely broken. The third phase is
characterized by the absence of atomic ODLRO but where molecules do
condense, the MSF
phase~\cite{kuklov04_1,kuklov04_2,Radzihovsky2004,Romans2004a,Radzihovsky2008a}. The
MSF phase has a residual $U(1)$ symmetry (rotations with
$\phi_\uparrow + \phi_\downarrow=\text{const}.$).

For $\delta \lesssim 1.6$~meV, Fig.~\ref{fig:phase} shows a second
order N-AMSF transition, as expected when bipolaritons are
irrelevant. As $\delta$ increases, $\nu$ decreases and the bare
resonance occurs at $\delta=3.84$~meV. The MSF phase appears at
slightly higher $\delta \simeq 4.0$~meV.  This is because near the
resonance --- between the lower two tricritical points (solid [blue]
circles) --- the the N-AMSF phase transition becomes first order,
leading to phase separation (right panel of Fig.~\ref{fig:phase}), a
dramatic signature of the resonance effects.
Phase separation occurs either between N and AMSF below the critical
end-point (solid [red] diamond at $\delta \simeq 4.0$~meV), or above,
between MSF and AMSF.  Above the second tricritical point, there is a
second order N-MSF transition at $2\mu=\nu$, which is then followed by
a second order MSF-AMSF transition. This MSF-AMSF transition occurs
when the renormalized bipolariton energy (due to the interaction
$\tilde{U}_X^{mm}$) reaches the polariton energy, as discussed
in~\cite{Basu:Stability}. At even higher detunings, the MSF-AMSF
transition becomes first order again, driven by the changing polariton
dispersion, (see~\cite{SM}).

As well as clear signatures in the form of the phase diagram, the
different collective paired phases can be experimentally identified
via spatial correlation functions. In particular, at $T=0$, the AMSF
phase is characterized by ODLRO of both unpaired polaritons and
bipolaritons. At $T\neq 0$, as the system is 2D, this evolves into
off-diagonal quasi long-range order (ODqLRO), i.e., power-law decay of
the correlation functions $g^{(1)}_{\uparrow,\downarrow} (\vect{r}) =
\langle \hat{\psi}_{\uparrow,\downarrow}^\dag(\vect{r})
\hat{\psi}_{\uparrow,\downarrow}^{}(0) \rangle$ and $g^{(1)}_m
(\vect{r}) = \langle \hat{\psi}_{\uparrow}^\dag(\vect{r})
\hat{\psi}_{\downarrow}^\dag(\vect{r}) \hat{\psi}_{\downarrow}^{}(0)
\hat{\psi}_{\uparrow}^{}(0) \rangle$. In contrast, the MSF phase is
characterized by the absence of any order for unpaired polaritons, but
the power-law decay of $g^{(1)}_m (\vect{r})$: The observation of such
pair correlations without polariton correlations would provide
unambiguous evidence for an MSF phase. An experimental scheme to
measure $g^{(1)}_m$ is given in~\cite{SM}.

To derive the $T=0$ phase diagram we employ a variational approach, by
considering a normalized Bogoliubov--Nozi\`eres ground
state~\cite{nozieres82} including atomic and molecular condensates, as
well as pairing terms:
\begin{multline*}
  |\psi \rangle = \mathcal{N} \exp \Bigg( \displaystyle
  \sqrt{\mathcal{A}} \sum_{\sigma= \uparrow, \downarrow, m}
  \psi_{\sigma} \hat{a}_{0\sigma}^\dag \\ + \sum_{\vect{k}}
  \sum_{\gamma=a, b, m} \tanh \theta_{\vect{k}\gamma}
  \hat{b}_{\vect{k}\gamma}^\dag \hat{b}_{-\vect{k}\gamma}^\dag\Bigg)
  |0\rangle\; .
\end{multline*}
The operators $\hat{b}_{\vect{k} \gamma}^{}$ are related to
$\hat{a}_{\vect{k} \sigma}^{}$ by:
\begin{align*}
  \begin{pmatrix}
  \hat{a}_{\vect{k}\uparrow}^{\dag} \\
  \hat{a}_{\vect{k}\downarrow}^\dag \end{pmatrix} & =
\Frac{1}{\sqrt{2}} \begin{pmatrix} 1 & 1 \\ -1 & 1 \end{pmatrix}
  \begin{pmatrix}
    \hat{b}_{\vect{k}a}^{\dag}
    \\ \hat{b}_{\vect{k}b}^\dag \end{pmatrix} &
    \hat{a}_{\vect{k}m}^{\dag}
    &=
    \hat{b}_{\vect{k}m}^{\dag} \;. 
\end{align*}
Note that $\sigma= \uparrow, \downarrow, m$, while $\gamma=a, b, m$.
This transformation produces the most general ($\uparrow
\leftrightarrow \downarrow$ symmetric) variational ground state
including pairing.
Minimizing the energy $\langle \hat{H} \rangle$ over the variational
parameters $\psi_\uparrow = \psi_\downarrow = \psi_0$, $\psi_m$, and
$\theta_{\vect{k}\gamma}$ we find that $\theta_{\vect{k}\gamma}$ has
the functional form $\tanh 2 \theta_{\vect{k}\gamma} =
{\alpha_\gamma}/{(\epsilon_{\vect{k}\gamma} + \beta_\gamma)}$, with
$\beta_\gamma>0, |\alpha_\gamma| \le \beta_\gamma$, and
$\epsilon_{\vect{k}a,b} = \omega_{\vect{k}}^{LP} - \omega_{0}^{LP}$.
The energy can thus be numerically minimized in terms of eight
variational parameters $\alpha_\gamma, \beta_\gamma$, and
$\psi_{\sigma}$, making it easy to determine first order phase
boundaries, as well as to find cases where the global minimum energy
is not an extremum (zero derivative), but instead occurs at
$|\alpha_\gamma| = \beta_\gamma$.

As $\tilde{U}^{\sigma \sigma'} \ll \tilde{U}^{\sigma \sigma'}_X \sim
1$ fluctuation corrections to mean field (MF) theory should be small.
At the same time, as bipolaritons have a much larger mass than LPs,
bipolariton fluctuations give a non-negligible shift. The dashed lines
and empty symbols in Fig.~\ref{fig:phase} show the MF predictions.
Fluctuations do shift the phase boundaries, but the phase diagram
topology qualitatively matches MF predictions~\cite{Radzihovsky2008a},
except for the TCP at large detuning where the MSF-AMSF transition
becomes first order again (see~\cite{SM}).  At MF level the two
tricritical points are at $(\mu, \nu) = (0,
\Delta^{LP}_{00}/(4\bar{U}))$ and $( -\Delta^{LP}_{00}/(2\bar{W}),
-\Delta^{LP}_{00}[\bar{W}^{-1} + (8 \bar{U})^{-1} ])$ where
$\bar{U}=\tilde{U}^{\uparrow \uparrow} + \tilde{U}^{\uparrow
  \downarrow}$ and $\bar{W}=[\tilde{U}^{mm} (\tilde{U}^{\uparrow
  \uparrow} + \tilde{U}^{\uparrow \downarrow})]^{1/2}$.

\paragraph{Finite temperature ---}
We have shown that, in GaAs, a polariton MSF phase can be found at
$T=0$ for $\delta \gtrsim 4.0$~meV.  However, to see if such a phase
is readily accessible, we must determine its critical
temperature. Since the closed channel mass is $m_M \simeq 2 m_X$, the
critical temperature is expected to be much lower than corresponding
LP condensation temperatures.

In order to extend our results to finite temperature, we use
variational mean-field theory (VMFT)~\cite{Radzihovsky2008a,kleinert},
based on the inequality~\cite{Feynman1998} $\mathcal{F} =-k_BT \ln \tr
e^{-\hat{H}/k_B T} \le \mathcal{F}_{v\MF} = \mathcal{F}_\MF + \langle
\hat{H} - \hat{H}_\MF \rangle_\MF$.  In a similar spirit to the $T=0$
calculation, $\hat{H}_\MF$ is chosen to allow the same one- and
two-point correlation functions and its variational parameters are
used to minimize $\mathcal{F}_{v\MF}$:
\begin{multline*}
  \hat{H}_\MF = \sum_{\gamma} \Bigg\{-\sqrt{\mathcal{A}} \psi_{\gamma}
  \left(\alpha_{\gamma}+ \beta_{\gamma} \right) \left(
  \hat{b}_{0\gamma}^\dag + \hat{b}_{0\gamma}^{}\right) \\ +
  \Frac{1}{2} \sum_{\vect{k}}\begin{pmatrix}
    \hat{b}_{\vect{k}\gamma}^\dag &
    \hat{b}_{-\vect{k}\gamma}^{} \end{pmatrix} \begin{pmatrix}
    \epsilon_{\vect{k}\gamma} + \beta_\gamma & \alpha_\gamma
    \\ \alpha_\gamma & \epsilon_{\vect{k}\gamma} +
    \beta_\gamma \end{pmatrix} \begin{pmatrix}
    \hat{b}_{\vect{k}\gamma}^{}
    \\ \hat{b}_{-\vect{k}\gamma}^\dag \end{pmatrix} \Bigg\} \; ,
\end{multline*}
where $\psi_a = 0$ and $\psi_b = \sqrt{2} \psi_0$. 
As before, the functional form of the matrix elements above is the
optimal form.
We evaluate $\mathcal{F}_\MF$ and averages $\langle \ldots
\rangle_{\text{MF}}$ by standard Bogoliubov
diagonalization~\cite{pitaevskii03}.  This yields the free energy
$\mathcal{F}_{v\MF}$ as a function of the same eight parameters
$\alpha_\gamma$, $\beta_\gamma$, and $\psi_0, \psi_m$, again allowing
numerical minimization.
Note that the $T=0$ limit of this approach reproduces the results
presented above.

\begin{figure}
\centering
\includegraphics[width=3.2in]{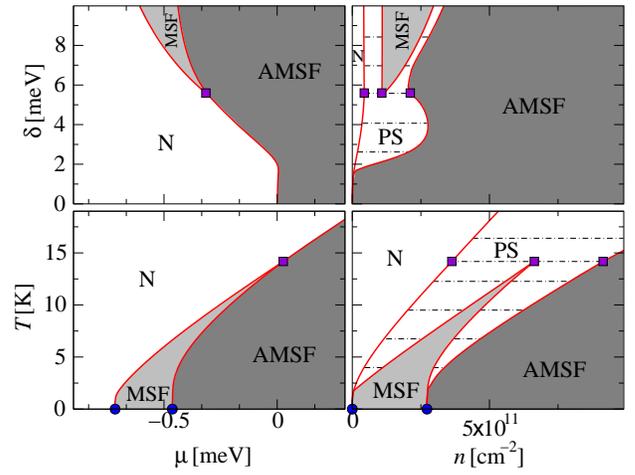}
\caption{(Color online) Finite temperature phase diagrams with the
  same notation, color scheme and GaAs parameters as in
  Fig.~\ref{fig:phase}. The critical end-point of Fig.~\ref{fig:phase}
  is now replaced by a triple point ([purple] square). Temperature is
  fixed to $T=4$~K ($k_BT= 0.34$~meV) in the top panels, while
  detuning is $\delta=10$~meV in the bottom ones.}
\label{fig:finit}
\end{figure}

For 2D Bose gases the actual $T \neq 0$ transition to a superfluid
phase is of the Berezinskii-Kosterlitz-Thouless (BKT)
type~\cite{pitaevskii03}.  Instead, VMFT predicts a first order
transition. The N-AMSF and MSF-AMSF transitions can be very weakly
first order due to the small polariton mass~\cite{SM}.  Despite the
absence of true ODLRO, the quasi-condensate density plays a similar
role to the mean-field order
parameter~\cite{kagan00,Prokofev2002}. This allows
Hartree-Fock~\cite{popov,pitaevskii03} or equivalent approaches (such
as VMFT) to reproduce the equation of state of the system outside of
the critical region.
As such, the location of phase boundaries predicted by VMFT should be
accurate, and we have verified this for the homonuclear weakly
interacting 2D Bose gas by comparison to the Monte-Carlo results
of~\cite{Prokofev2002}.
In the 3D homonuclear case, other approaches such as applications of
the Nozi\`eres-Schmitt-Rink approximation~\cite{Nozieres1985} have
been considered, predicting  only  second order
transitions~\cite{Koetsier2009}.

Figure~\ref{fig:finit} shows the phase diagram both vs $\delta$ at
fixed $T$ (top panels) and vs $T$ at fixed $\delta$ (bottom panels).
The main new features introduced by non-zero temperature are the
existence of a non-zero critical density for the normal state and the
extension to arbitrarily large detuning of the phase separated region
as discussed above. Because the N-MSF transition is now first order,
the critical end-point at $T=0$ is replaced by a triple point.
At yet higher temperatures, the triple point moves to higher $\delta$
and eventually merges with the tricritical point (see~\cite{SM}).
In the phase diagram vs temperature at $\delta=10$~meV, one sees that
the MSF phase survives up to $T \sim 14.2$~K ($1.28$~meV).  For
ZnO~\cite{Li2013a} the MSF phase can survives to higher temperatures
and lower detunings~\cite{SM}.

\paragraph{Conclusions}
To conclude, we argue that microcavity polaritons are an ideal system
to explore collective pairing phases of bosons. In particular, in
addition to the standard polariton superfluid phase, where both
polaritons and bipolaritons are characterized by off-diagonal quasi
long-range order, we highlight a new phase which displays molecular
superfluidity (MSF), i.e., order for bipolaritons but not for
polaritons. This phase covers an increasing region of the phase
diagram at either larger cavity-exciton detunings or lower
temperatures. While for the GaAs parameters considered here we predict
the MSF phase to survive up to $T\sim 14$~K at a cavity-exciton
detuning $\delta=10$~meV, this temperature can be higher for materials
with larger Rabi splitting, such as ZnO.

\acknowledgments We are grateful to B. D. Simons, L. Radzihovsky, and
M. Wouters for stimulating discussions.
FMM acknowledges financial support from the programs Ram\'on y Cajal,
MINECO (MAT2011-22997), CAM (S-2009/ESP-1503), and Intelbiomat (ESF).
JK acknowledges support from EPSRC programme ``TOPNES'' (EP/I031014/1)
and EPSRC (EP/G004714/2).


\newpage
\setcounter{page}{1}

\pagenumbering{gobble}

\setcounter{secnumdepth}{2}

\renewcommand{\figurename}{\textsc{S.~Fig.}}

\setcounter{equation}{0}
\setcounter{figure}{0}

\onecolumngrid
\begin{center}
\large
\textbf{Supplemental Material for ``Collective pairing of resonantly
  coupled microcavity polaritons''}  
\end{center}
\vspace{3ex}
\twocolumngrid

\section{Model parameters}
\label{sec:detun}
Our manuscript calculates the phase diagram of the polariton system
using the two-channel model, given in Eq.~(1) of the manuscript.  This
model describes the interacting lower polariton (LP) fields in
  the two circular polarization states
  $\hat{\psi}_{\uparrow,\downarrow} (\vect{r}) = [\hat{\psi}_x
    (\vect{r}) \pm i \hat{\psi}_y (\vect{r})]/\sqrt{2}$ ($x,y$ are the
  linear polarization components) and the bipolariton $\hat{\psi}_m
(\vect{r})$ field. Such a model is the low-energy effective theory of
a more complicated model of interacting excitons strongly coupled to
photons.  Most properties of the bipolariton field in our model are
the same as those of pure biexcitons, although the bipolariton does
contain a small photon component. This small photon component provides
crucial distinctions from a pure biexciton state.  In particular, the
photon component ensures that optically active states are the lowest
in energy.

To clarify these points, we give here a short account of the
derivation of the effective polariton and bipolariton parameters
appearing in the two-channel model, such as the effective interaction
strengths $U^{\sigma \sigma'}_{\vect{k}\vect{k^\prime}\vect{q}}$
(where $\sigma, \sigma'= \uparrow, \downarrow, m$) and the Feshbach
resonance coupling $g_{\vect{k}\vect{Q}}$, the inter-channel detuning
$\nu$, the bipolariton mass $m_m$, and their dependence on the
exciton-photon detuning $\delta = \omega_0^C - \omega_0^X$, where
$\omega_{\vect{k}}^{C,X}$ are the photon and exciton
energies. Changing the detuning $\delta$ changes the LP composition,
and hence affects the polariton dispersion, its scattering properties,
and the overlap with the biexciton states. Many of these results are
also reported in the manuscript, for completeness we include all
relevant formulae here.

The LP dispersion $\omega_{\vect{k}}^{LP}$ has the form $2
\omega_{\vect{k}}^{LP} = \omega_{\vect{k}}^{C} + \omega_{\vect{k}}^{X}
- \sqrt{(\omega_{\vect{k}}^{C} - \omega_{\vect{k}}^{X})^2 +
  \Omega_R^2}$, as given in (and shown in Fig. 1 of) the manuscript.
The non-quadratic form of this dispersion results from the varying
composition of the LP, as determined by the Hopfield coefficients,
$2c_{\vect{k}}^2, 2s_{\vect{k}}^2 = 1 \pm (\omega_{\vect{k}}^{C} -
\omega_{\vect{k}}^{X})/\sqrt{(\omega_{\vect{k}}^{C} -
  \omega_{\vect{k}}^{X})^2 + \Omega_R^2}$, which give the exciton and
photon fraction of the lower polariton at a given momentum $\vect{k}$.

In order to derive the appropriate dimensionless parameters
characterizing the interaction strengths, it is necessary to
identify an effective mass of the LP. This corresponds to the
expansion at small $\vect{k}$, $\omega_{\vect{k}}^{LP} -
  \omega_{0}^{LP} \simeq k^2/2m$, where $m = m_{LP} (\vect{k}=0)$.
This polariton mass is given by
\begin{equation}
  \frac{m}{m_X} = \frac{1}{c_{0}^2 + s_{0}^2 m_X/m_C}\; ,
\end{equation}
where the Hopfield coefficients at $\vect{k}=0$ take a simple form:
\begin{equation}
  c_{0}^2, s_{0}^2= \Frac{1}{2} \left[1 \pm
    \Frac{\delta}{\sqrt{\delta^2 + \Omega_R^2}}\right].
\label{eq:Hopf0}  
\end{equation}
We will discuss what changes would occur if this quadratic
approximation of the LP dispersion were used in calculating the phase
diagram; in particular, in Sec.~\ref{sec:quadratic-vs-non}, we will
compare these results with those obtained in the manuscript by making
use of the full LP dispersion.

From the LP dispersion, we can also express the detuning between
closed and open channels (the inter-channel detuning), $\nu= -|E_b| -
2 [\omega_0^{LP} - \omega_0^X]$, in terms of the exciton-photon
detuning $\delta$:
\begin{equation}
  \nu = - |E_b| - \delta + \sqrt{\delta^2 + \Omega_R^2} \; .
\end{equation}
A plot of the dependence of $\nu$ and $m/m_X$ on the detuning $\delta$
for typical GaAs parameters can be found in the manuscript Fig.~1(c).

The background polariton interaction strengths $U^{\uparrow \uparrow}
= U^{\downarrow \downarrow}$, $2 U^{\uparrow \downarrow}$ can be
written in terms of the corresponding excitonic interaction
strengths. They depend on the polariton or exciton mass used to define
dimensionless interaction constants $\tilde{U}^{\sigma \sigma'}$ and
$\tilde{U}^{\sigma \sigma'}_X$, as well as on the excitonic fraction,
i.e., the Hopfield coefficient $c_{\vect{k}}$, which introduces a
non-trivial dependence of the polariton interaction strengths on the
momentum ($\hbar = 1$):
\begin{equation}
  U_{\vect{k} \vect{k}' \vect{q}}^{\sigma \sigma'} =
  \frac{\tilde{U}_{\vect{k} \vect{k}' \vect{q}}^{\sigma \sigma'}}{m} =
  c_{\vect{k}} c_{\vect{k}'} c_{\vect{k}'-\frac{\vect{q}}{2}}
  c_{\vect{k} + \frac{\vect{q}}{2}} \frac{\tilde{U}^{\sigma
      \sigma'}_X}{m_X} \; ,
\end{equation}
where $\sigma, \sigma^\prime$ run over $\uparrow, \downarrow$ only
here. The background bipolariton interaction $U^{mm}$ does not involve
the polariton Hopfield coefficients because, as we will see below, the
bipolariton is strongly excitonic regardless of momentum.

The hybridization (coupling) $g_{\vect{k}\vect{Q}}$ between pairs of
LPs in opposite polarizations and bipolaritons, as well as the
bipolariton mass $m_m$, can be estimated from the expression for the
LP-LP scattering $T$-matrix, $t_{LPLP}(E, \vect{k},\vect{k}^\prime,
\vect{Q})$. The bipolariton appears as a scattering resonance in this
$T$-matrix, and its mass depends on the dispersion of that
resonance. The $T$-matrix can be evaluated starting from the
$T$-matrix for two excitons in opposite polarizations, $t_{XX}
(E,\vect{k},\vect{k}^\prime, \vect{Q})$, scattering from momenta
$(\vect{k}+\vect{Q}/2, -\vect{k}+\vect{Q}/2)$ to momenta
$(\vect{k}^\prime+\vect{Q}/2,
-\vect{k}^\prime+\vect{Q}/2)$. Assuming that the exciton
  scattering is dominated by the biexciton resonance, one can assume
  the following simplified expression for $t_{XX}
  (E,\vect{k},\vect{k}^\prime, \vect{Q})$:
\begin{equation}
  t_{XX} (E,\vect{k},\vect{k}^\prime, \vect{Q}) \simeq
  \frac{4\pi}{m_X} \Frac{\Delta^X}{E - \omega_{XX} - \frac{Q^2}{2
      (2m_X)}}\; .
\end{equation}
We then include the exciton-photon coupling to evaluate $t_{LPLP}(E,
\vect{k},\vect{k}^\prime, \vect{Q})$. Using the results of
Ref.~\cite{Wouters2007}, one has that
\begin{equation}
  t_{LPLP}(E, \vect{k},\vect{k}^\prime, \vect{Q})=\Frac{c_{\vect{k} +
      \frac{\vect{Q}}{2}} c_{-\vect{k} + \frac{\vect{Q}}{2}}
    c_{\vect{k}^\prime + \frac{\vect{Q}}{2}} c_{-\vect{k}^\prime +
      \frac{\vect{Q}}{2}} }{t_{XX}^{-1}(E , \vect{k},\vect{k}^\prime,
    \vect{Q})- D(E ,\vect{Q})} \; ,
\end{equation}
where
\begin{multline}
  D(E ,\vect{Q}) = \sum_{\vect{p}} \left[\Frac{c_{\vect{p}}^2
      c_{-\vect{p} + \vect{Q}}^2}{E-\omega_{\vect{p}}^{LP} -
      \omega_{-\vect{p} + \vect{Q}}^{LP}} - \Frac{1}{E- 2
      \omega_{\vect{p}}^{X}} \right. \\ \left. + \Frac{2
      c_{\vect{p}}^2 s_{-\vect{p} +
        \vect{Q}}^2}{E-\omega_{\vect{p}}^{LP} - \omega_{-\vect{p} +
        \vect{Q}}^{UP}} + \Frac{s_{\vect{p}}^2 s_{-\vect{p} +
        \vect{Q}}^2}{E- \omega_{\vect{p}}^{UP} - \omega_{-\vect{p} +
        \vect{Q}}^{UP}}\right]\; ,
\end{multline}
and where the dispersion of the upper polariton (UP) is $2
\omega_{\vect{k}}^{UP} = \omega_{\vect{k}}^{C} + \omega_{\vect{k}}^{X}
+ \sqrt{(\omega_{\vect{k}}^{C} - \omega_{\vect{k}}^{X})^2 +
  \Omega_R^2}$.  The propagator $D(E,\vect{Q})$ is generally a smooth
function of the energy $E$, except at isolated points.  For zero
center of mass momentum, $\vect{Q}=0$, these isolated points, where it
diverges logarithmically, occur near the 2-particle band-edge energies
(i.e., $E \sim 2 \omega_{0}^{LP}$, $E \sim \omega_{0}^{LP} +
\omega_{0}^{UP}$, and $E \sim 2 \omega_{0}^{UP}$). Because $m_C \ll
m_X$, the ratio between polariton and exciton mass $m/m_X$ also
generally satisfies $m/m_X \ll 1$, unless at very large detuning
$\delta \gg \Omega_R$ (see manuscript Fig.~1(c)).  The existence of
this small parameter suppresses the renormalization term $D(E
,\vect{Q})$.  In general it causes only a small shift of the resonance
energy away from $\omega_{XX} + \frac{Q^2}{2 (2m_X)}$, thus giving:
\begin{align}
  t_{LPLP}(E, \vect{k},\vect{k}^\prime,\vect{Q}) &\simeq
  \frac{4\pi}{m} \Frac{ g_{\vect{k}\vect{Q}}
    g_{\vect{k}^\prime\vect{Q}} m}{E - \omega_{XX} - \frac{Q^2}{2
      (2m_X)}},\\ g_{\vect{k}\vect{Q}} &= c_{\vect{k} +
    \frac{\vect{Q}}{2}} c_{-\vect{k} + \frac{\vect{Q}}{2}}
  \sqrt{\Delta^X/m_X}.
\end{align}
The Feshbach coupling $g_{\vect{k}\vect{Q}}$ is related to the
resonance width $\Delta^{LP}_{\vect{k}\vect{Q}}$ by
$\Delta^{LP}_{\vect{k}\vect{Q}}=m
g_{\vect{k}\vect{Q}}^2$~\cite{Gurarie2007}, so that
\begin{equation}
  \Delta^{LP}_{\vect{k}\vect{Q}} = c_{\vect{k} + \frac{\vect{Q}}{2}}^2
  c_{-\vect{k} + \frac{\vect{Q}}{2}}^2 \Delta^X \Frac{m}{m_X} \; .
\end{equation}
We have verified that, for the system parameters relevant for current
experiments in GaAs (see later section ``GaAs parameters''), the above
approximation,
\begin{equation}
  g^2_{\vect{k}\vect{Q}} \simeq c_{\vect{k} + \frac{\vect{Q}}{2}}^2
  c_{-\vect{k} + \frac{\vect{Q}}{2}}^2 \Delta^X/m_X\; ,
\end{equation}
is accurate up to $\delta \lesssim 33$~meV, and $0.1$~meV away from
the resonance (i.e., $\nu = 0$ which corresponds to
$\delta=3.84$~meV), with an error below $5\%$, in the limit $\vect{k}
\to 0$ and $\vect{Q} \to 0$.
%

Thus we have seen that most properties of bipolaritons are the same as
those of pure biexcitons. However, even when the photon fraction is
very small, the photon component has crucial effects. Firstly, given
the small ratio, $m_C/m_X = 10^{-4}$, even a 1\% photon fraction
reduces the polariton mass $m$ by two orders of magnitude compared to
the bare excitonic mass.  As such, a 99\% excitonic polariton system
can behave quite distinctly from a pure excitonic
system~\cite{Li2013a,Trichet2013}. In addition to the change of mass
(relevant for when later we will make use of the quadratic
  approximation of the LP dispersion) there is another significant
effect on excitons when including a small mixing with photons. In
GaAs, excitons can have angular momentum $J_z=\pm 2, \pm 1$.  These
states are almost degenerate, with a weak (typically a few
$\mu$eV~\cite{VanKesteren1990a}) electron-hole exchange term favoring
the ``dark'' $J_z = \pm 2$ states~\cite{Combescot2007a}. Photons
couple only to the optically active ``bright'' state $J_z=\pm 1$ \and
thus lower the energy of these states.  If the Rabi splitting is
typically $10^3$ times larger than the exchange splitting, i.e. a few
meV rather than a few $\mu$eV, then only 0.1\% photon fraction is
required to overcome the exchange splitting.  Thus, while the ground
state of a pure excitonic system is expected to be dark, the ground
state of 99.9\% excitonic system is likely
bright~\cite{Li2013a,Trichet2013}. Such photon induced splitting of
bright and dark states can also be important for the bipolariton, even
when the bipolariton mass and energy are very similar to those of the
biexciton.

\begin{figure}
\centering
\includegraphics[width=3.2in]{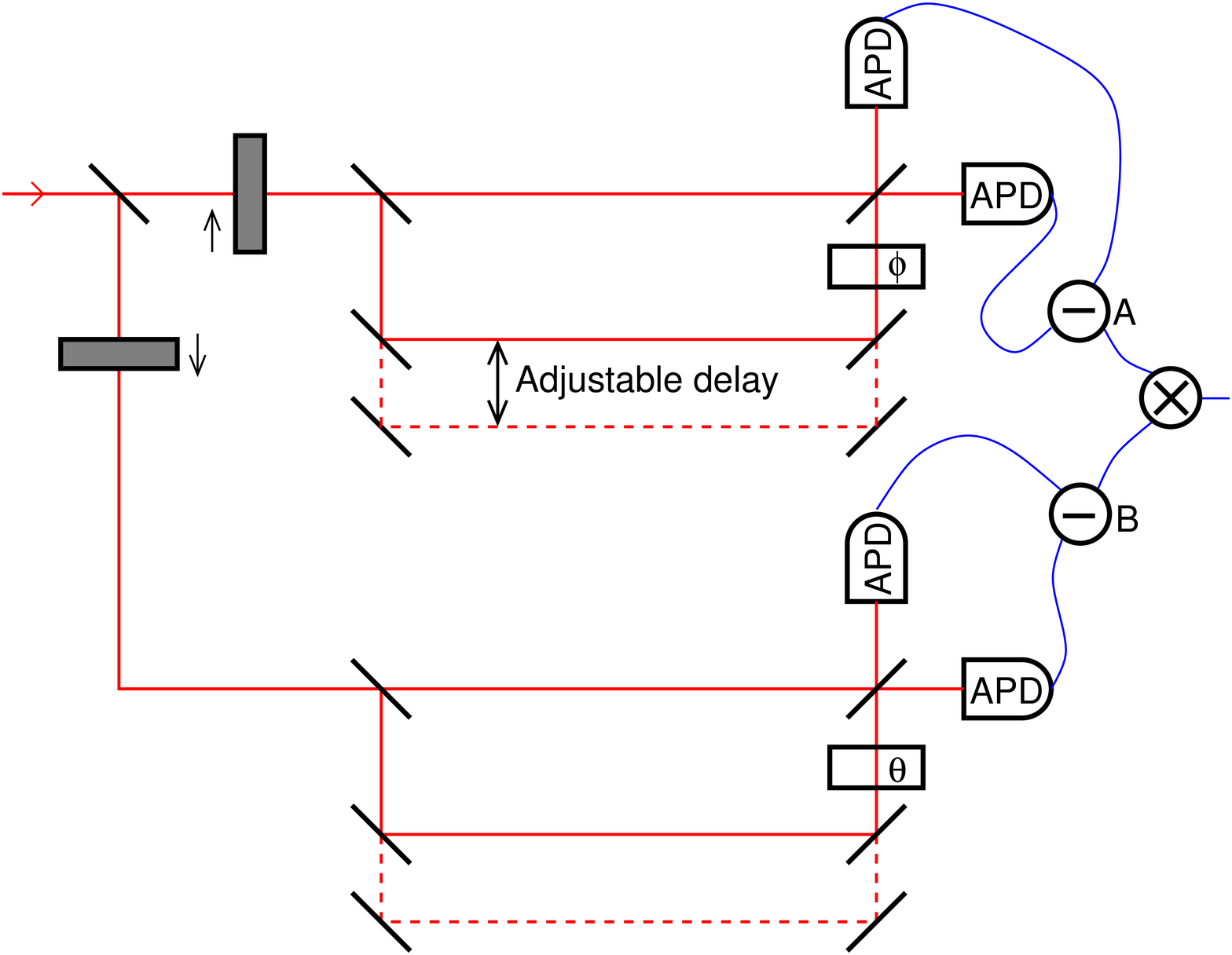}
\caption{Schematic optical detection setup for measuring
  time-dependent pair correlations $g^{(1)}_m(0,t)$.}
\label{fig:schem-expt}
\end{figure}
%
\section{Detection of pairing correlations}
\label{sec:detect-pair-corr}
In the manuscript we discussed the distinct signatures for detecting
the MSF and AMSF phases. The clearest signature is that the MSF phase
would show pair coherence, but no polariton coherence.  Experiments to
determine whether polariton coherence exists, by measuring
$g^{(1)}_\sigma(\vect{r}, t)=\langle
\hat{\psi}_{\sigma}^\dag(\vect{r}, t) \hat{\psi}_{\sigma}^{}(0,0)
\rangle$ are routine and can be extracted from interference fringe
visibility maps between two images of the condensate. Measuring such
correlations clearly distinguishes the AMSF phase from the MSF or
Normal phase. However, to unambiguously distinguish between the normal
and the MSF phases it is necessary instead to measure the following
molecular first order correlation function $g^{(1)}_m (\vect{r}, t) =
\langle \hat{\psi}_{\uparrow}^\dag(\vect{r}, t)
\hat{\psi}_{\downarrow}^\dag(\vect{r}, t)
\hat{\psi}_{\downarrow}^{}(0, 0) \hat{\psi}_{\uparrow}^{}(0, 0)
\rangle$.  In this section we present one possible scheme to measure
$g^{(1)}_m(0,t)$; a similar setup with spatial resolution could be
used to measure $g^{(1)}_m(\vect{r},t)$. Other possible schemes to
measure these quantities are possible, our aim here is to illustrate
the possibility of such measurements.

Supplemental Figure~\ref{fig:schem-expt} shows a possible experimental
scheme that measures the pair correlations $g^{(1)}_m(0,t)$. The
scheme is based on cross-correlation measurements between a pair of
balanced homodyne detectors, which separately measure two-time
correlations of the different polarization states.  Light from the
sample is split on an initial beamsplitter and filtered into the two
circular polarization components, $\uparrow$ and $\downarrow$.  This
light is then sent through a variable delay interferometer, with an
adjustable phase shift $\phi$ for $\uparrow$ polarization ($\theta$
for $\downarrow$ polarization). The same delay is used for both the
$\uparrow$ and $\downarrow$ branches of the apparatus. The light is
then measured via a balanced homodyne scheme, so that at $A$ the
output is:
\begin{align*}
  I_A &= \frac{1}{2} \langle [\hat{\psi}^\dagger_{\uparrow} (t) -i
    e^{-i \phi} \hat{\psi}^\dagger_{\uparrow} (t') ][
    \hat{\psi}^{}_{\uparrow} (t) + i e^{i \phi}
    \hat{\psi}^{}_{\uparrow}(t') ]\rangle \\ &- \frac{1}{2} \langle
  [-i\hat{\psi}^\dagger_{\uparrow} (t) + e^{-i \phi}
    \hat{\psi}^\dagger_{\uparrow} (t') ][ i \hat{\psi}^{}_{\uparrow}
    (t) + e^{i \phi} \hat{\psi}^{}_{\uparrow}(t') ]\rangle\\ &= i e^{i
    \phi} \langle \hat{\psi}^\dagger_{\uparrow} (t)
  \hat{\psi}^{}_{\uparrow}(t')\rangle + \text{h.c.}\; ,
\end{align*}
where $t'=t+\tau$, $\tau$ is the adjustable delay, and $\phi$ the
phase shift. This signal on its own averages to zero in the MSF phase
as there is no single polariton phase coherence. A similar form
applies for the lower branch of the interferometer $B$ with $\uparrow
\mapsto \downarrow$ and $\phi \mapsto \theta$. When calculating the
cross correlation $I_{\text{out}} = \langle I_A \times I_B \rangle$,
one therefore obtains a signal:
\begin{align}
  I_{\text{out}} =& - e^{i (\phi + \theta)} \langle
  \hat{\psi}^\dagger_\uparrow(t) \hat{\psi}^{}_\uparrow(t')
  \hat{\psi}^\dagger_\downarrow(t) \hat{\psi}^{}_\downarrow(t')
  \rangle \nonumber\\&+ e^{i(\phi - \theta)} \langle
  \hat{\psi}^\dagger_\uparrow(t) \hat{\psi}^{}_\uparrow(t')
  \hat{\psi}^\dagger_\downarrow(t') \hat{\psi}^{}_\downarrow(t)
  \rangle \nonumber\\&+ e^{i(\theta - \phi)} \langle
  \hat{\psi}^\dagger_\uparrow(t') \hat{\psi}^{}_\uparrow(t)
  \hat{\psi}^\dagger_\downarrow(t) \hat{\psi}^{}_\downarrow(t')
  \rangle \nonumber\\& - e^{-i (\phi + \theta)} \langle
  \hat{\psi}^\dagger_\uparrow(t') \hat{\psi}^{}_\uparrow(t)
  \hat{\psi}^\dagger_\downarrow(t') \hat{\psi}^{}_\downarrow(t)
  \rangle \; .
\end{align}
In the absence of a symmetry breaking magnetic field, correlations
should be invariant under the replacement $\uparrow \leftrightarrow
\downarrow$ and so the middle two lines involve the same expectation.
Choosing $\phi-\theta = \pi/2$ then means these middle two lines
vanish.  Choosing also $\phi+\theta=\pi$, so that $\phi = 3\pi/4,
\theta=\pi/4$, the first and last lines have the same prefactor
giving:
\begin{align}
  I_{\text{out}} &= \langle \hat{\psi}^\dagger_\uparrow(t)
  \hat{\psi}^{}_\uparrow(t') \hat{\psi}^\dagger_\downarrow(t)
  \hat{\psi}^{}_\downarrow(t') \rangle \nonumber\\&+ \langle
  \hat{\psi}^\dagger_\uparrow(t') \hat{\psi}^{}_\uparrow(t)
  \hat{\psi}^\dagger_\downarrow(t') \hat{\psi}^{}_\downarrow(t)
  \rangle\; .
\end{align}
In the steady state, this implies that $I_{\text{out}} \propto
g_m^{(1)}(0,\tau) + g_m^{(1)}(0,-\tau)$, and as the coherent part of
the two-time correlation function should be symmetric in $\tau$, such
a scheme allows measurement of the pair coherence.

\section{Comparing the case of full LP dispersion 
with the quadratic approximation limit}
\label{sec:quadratic-vs-non}
In the manuscript we always use the full LP dispersion
$\omega^{LP}_{\vect{k}}$.  It is however instructive to compare this
to the quadratic approximation $\omega_{\vect{k}}^{LP} -
\omega_{0}^{LP} \simeq k^2/2m$ discussed above. This approximation
corresponds to assuming that the occupation of the low-momentum
  polariton states dominates all relevant expectations values, and so
in this limit it is consistent to also replace all Hopfield factors
with $c_{\vect{k}}, s_{\vect{k}} \to c_0, s_0$.

Approximating the LP dispersion by its low-momentum expansion is
expected to be accurate in the limit when fluctuations beyond
mean-field are small: The mean-field results do not depend on the
polariton dispersion at all, since they involve only the $\vect{k}=0$
fields.  As discussed in the manuscript, in the photonic limit the
small polariton mass $m \ll m_X$ suppresses fluctuations, however at
large $\delta$, fluctuations are significant, and the full dispersion
yields different predictions. Nonetheless, all the qualitative
features of the phase diagrams obtained by using the full LP
dispersion can be explained also by considering the quadratic
approximation.
  
\begin{figure}
\centering
\includegraphics[width=3.2in]{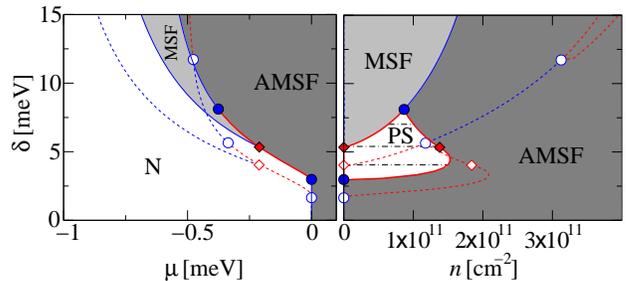}
\caption{Comparison between the phase diagrams at $T=0$ in the
  quadratic approximation $\omega_{\vect{k}}^{LP} - \omega_{0}^{LP}
  \simeq k^2/2m$ (solid lines), and with the full LP dispersion
  $\omega_{\vect{k}}^{LP}$ (dashed lines). The latter correspond to
  the boundaries shown in the Fig.~2 of the manuscript. Notation and
  color scheme are the same as in the manuscript.}
\label{fig:compare-q-nq}
\end{figure}

\begin{figure}
\centering
\includegraphics[width=3.2in]{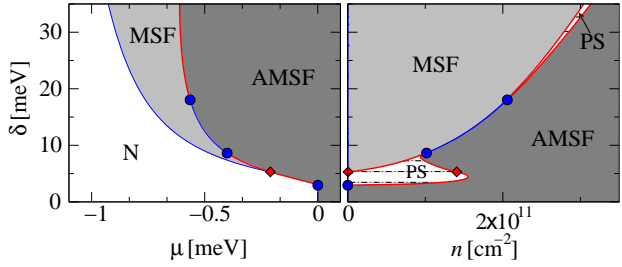}
\caption{Zero temperature phase diagram calculated in the quadratic
  approximation, but choosing an artificially large cavity photon mass
  $m_C/m_X = 10^{-3}$.  This shows that the phase separation
  occurring at large $\delta$ is not specific to the non-quadratic
  polariton dispersion.}
\label{fig:heavyt0}
\end{figure}

We plot in S.~Fig.~\ref{fig:compare-q-nq} the comparison between the
phase diagram at $T=0$ in the quadratic approximation, and with the
full dispersion. The basic topology is qualitatively the same,
  with the difference that in the quadratic approximation the MSF-AMSF
  phase separation re-emerges only at very large $\delta$ compared
  with the case of the full LP dispersion. This can also be seen more
clearly in S.~Fig.~\ref{fig:heavyt0} which shows the results of the
quadratic approximation, but choosing a ``heavier'' cavity photon
mass, $m_C=10^{-3} m_X$.  This demonstrates that this feature is not
dependent on the non-quadratic dispersion, but arises whenever the
polariton density of states becomes very large, as occurs at large
detunings. This in turn leads to a fluctuation dominated regime, which
can drive the phase boundary first order. Since the full LP dispersion
becomes strongly excitonic at large momentum, this generally increases
the polariton density of states as compared to the quadratic
approximation, enhancing such an effect.

\begin{figure}
\centering
\includegraphics[width=3.2in]{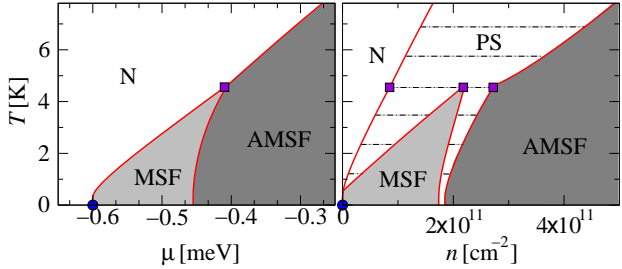}
\caption{Finite temperature phase diagram at fixed exciton-photon
  detuning $\delta=10$~meV, calculated in the quadratic approximation
  but taking artificially large cavity photon mass $m_C/m_X =
  10^{-2}$.  This is qualitatively similar to the behavior seen in
  Fig. 3 of the manuscript.}
\label{fig:heavydelta10}
\end{figure}

In a similar way, other apparent differences between the quadratic
approximation and the full dispersion cases arising at finite
temperature can also be shown to be reproduced in the quadratic
approximation with heavier photons. For example, the phase diagram at
$\delta=10$K shown in Fig. 3 of the manuscript, and
S.~Fig.~\ref{fig:gaas-vs-T} appears to behave differently at low
temperature.  The behavior of the full dispersion is however closely
replicated by the quadratic approximation if one takes $m_C/m_X =
10^{-2}$, as shown in S.~Fig.~\ref{fig:heavydelta10}.  This confirms
the statement that while the full dispersion introduces important
quantitative changes, no qualitatively new behavior occurs due to the
full dispersion.

The use of the quadratic approximation has the advantage to
  allowing
significantly faster determination of the phase boundaries in our
variational approach. The variational approach requires finding the
free energy for each value of the variational parameters, and this
requires estimating the integrals $\int d \vect{k} \langle
  a^\dagger_{\vect{k} \sigma} a^{}_{\vect{k} \sigma^\prime} \rangle$
  and $\int d \vect{k} \langle a^{}_{\vect{k} \sigma} a^{}_{\vect{k}
    \sigma^\prime} \rangle$. Within the quadratic approximation,
these integrals are almost all analytic. This significant speedup
allows a more comprehensive study of the three dimensional phase
diagram (temperature $T$, detuning $\delta$ versus either density $n$ or
detuning $\mu$), for which the full dispersion would be prohibitively
computationally expensive. As such, the remainder of this
supplemental material presents a comprehensive set of phase diagrams
calculated in the quadratic approximation.

\section{Quadratic approximation: Evolution of phase diagram 
with temperature and detuning }
\label{sec:evol-phase-diagr}
In the manuscript we presented several cuts through the
three-dimensional phase diagram as a function of temperature, detuning
and density (or chemical potential) for parameters relevant to GaAs.
These cuts illustrate the main features that can be seen, namely the
sequence of N-MSF-AMSF transitions and the direct N-AMSF transition;
the existence of a first order transition and the associated phase
separation (PS) near resonance at all temperatures, and the
temperature dependence of the various phase boundaries. In this
section we provide further cuts to illustrate more fully how the shape
of the phase diagram evolves with changing detuning and temperature.
Below we first present further figures illustrating the behavior for
the system parameters which are relevant for GaAs microcavities. Then
we compare the form of the phase diagram for GaAs parameters with that
for ZnO parameters.

\begin{figure}
\centering
\includegraphics[width=3.2in]{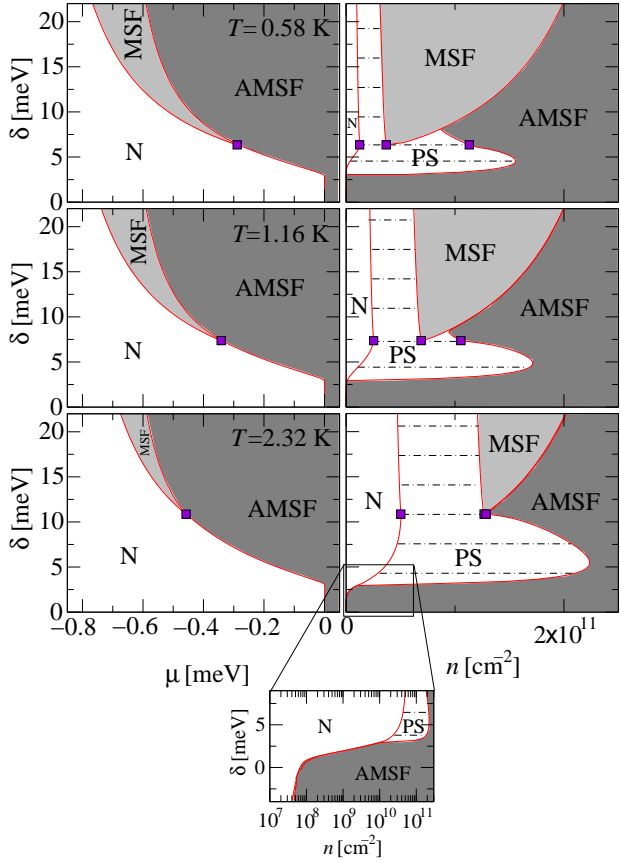}
\caption{Finite temperature phase diagrams for GaAs parameters in
    the low-momentum quadratic approximation for the LP dispersion
  plotted as a function of detuning $\delta$ for three different fixed
  temperatures ($T=0.58$~K, $T=1.16$~K, and $T=2.32$~K). Notation and
  color scheme are the same as those in the manuscript Fig.~3. For the
  last panel, corresponding to $T=2.32$~K, we also show the behavior
  of the N-PS-AMSF weakly first order boundary at low density and for
  negative detuning $\delta$ (please note the log-scale in
  density). As also discussed below, the behavior of the N-PS-AMSF
  first order transition for large and negative detunings $\delta$
  follows the BKT scaling $T \sim n/m$, where $m$ is the LP
  mass. Thus, at fixed temperature, it exists a small, but finite,
  critical density for the N-PS-AMSF transition which is around $4$
  order of magnitudes smaller than the critical density for the
  N-PS-MSF transition at positive and large detunings (which follows
  $T \sim n/m_m$).}
\label{fig:gaas-vs-detun}
\end{figure}
%
\subsection{GaAs parameters}
\label{sec:gaas-parameters}
\begin{figure}
  \centering
  \includegraphics[width=3.2in]{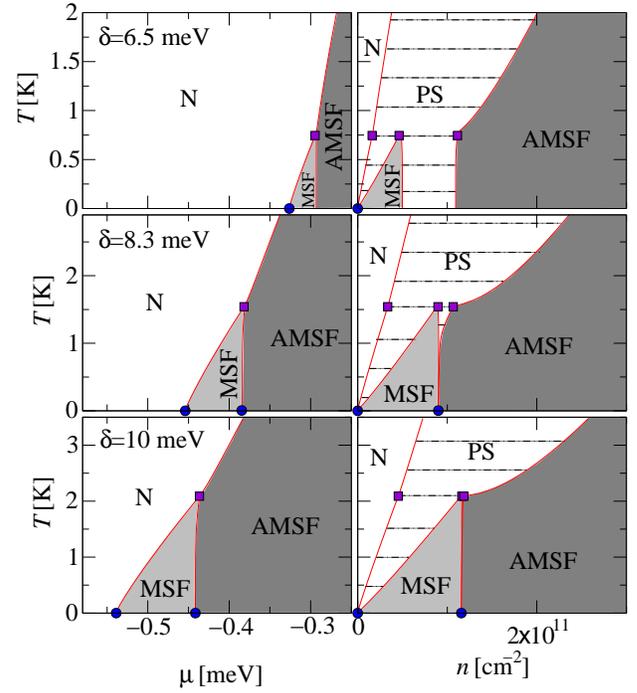}
  \caption{Finite temperature phase diagrams for GaAs parameters
    in the low-momentum quadratic approximation for the LP
      dispersion plotted as a function of temperature $T$ for three
    different fixed exciton-photon detunings ($\delta = 6.5$~meV,
    $\delta=8.3$~meV, and $\delta=10$~meV). Notation and color scheme
    are the same as the manuscript Fig.~3.}
\label{fig:gaas-vs-T}
\end{figure}

For GaAs, we fix the photon mass to $m_C = 10^{-4} m_X$, the exciton
mass $m_X = 0.4 m_e$, the Rabi splitting $\Omega_R=4.4$~meV, the
biexciton binding energy $|E_b|=2$~meV, and the excitonic resonance
width $\Delta^X=4|E_b|$~\cite{Wouters2007}. Additionally,
$\tilde{U}^{\uparrow \uparrow}_X = \tilde{U}^{\downarrow \downarrow}_X
= 6$~\cite{Wouters2007} and we also fix $\tilde{U}^{\uparrow
  \downarrow}_X = 0$ and $\tilde{U}^{mm}_X = 4$.
Additional cuts through the phase diagram to those provided in the
manuscript are here given in S.~Figs.~\ref{fig:gaas-vs-detun}
and~\ref{fig:gaas-vs-T}.  These show cuts at various fixed
temperatures (S.~Fig.~\ref{fig:gaas-vs-detun}) or fixed detunings
(S.~Fig.~\ref{fig:gaas-vs-T}).

The zero temperature phase diagram (see manuscript Fig.~2) is
characterized by phase boundaries that are mostly second order, with a
limited region in detuning near the resonance where the transition
either between MSF-AMSF or N-AMSF is first order, thus implying phase
separation (PS) for a range of densities. However, as
S.~Fig.~\ref{fig:gaas-vs-detun} and S.~Fig.~\ref{fig:gaas-vs-T}
clearly show, at any non-zero temperature, all phase boundaries that
were second order at zero temperature, evolve at finite temperature to
weakly first order --- as calculated within variational mean-field
theory (VMFT).
This is because, as discussed in the manuscript, at finite
temperature, the real phase transitions are of the
Berezinskii-Kosterlitz-Thouless type, involving a jump in the
superfluid density, and VMFT approximates this by a weakly first order
transition. 
The size of the jump depends on the mass of the condensing boson, and
so the N-AMSF or MSF-AMSF transitions (determined by the polariton
mass) show a very small jump. This density jump at the MSF-AMSF is
just visible in the cut at $T = 2.32$~K in
S.~Fig.~\ref{fig:gaas-vs-detun}, and is more pronounced in the
manuscript Fig.~3 top panel where $T=4$~K.  In the lower temperature
cuts in S.~Fig.~\ref{fig:gaas-vs-detun} the jump is hardly visible. In
contrast the jump at the N-MSF transition controlled by the biexciton
mass $m_m=2m_X$, and is hence much larger, and so is clearly visible
at all temperature in S.~Fig.~\ref{fig:gaas-vs-detun}.
Note also that, at large and negative detuning, the N-AMSF transition
follows the BKT scaling $T \sim n/m$, where $m$ is the LP mass, while
at large and positive detuning, the N-MSF transition follows $T \sim
n/m_m$ and it depends on the biexciton mass $m_m=2m_X$. In particular,
the AMSF region at large and negative detuning extends down up to a
critical finite density (at finite temperature), as it is clearly
shown in the low-density magnification of the bottom panel ($T =
2.32$~K) of S.~Fig.~\ref{fig:gaas-vs-detun}.

Since all phase boundaries are (at least weakly) first order, there
are no true tricritical points, however the cuts at $T=0.58$~K and
$T=1.16$~K show a remnant of the $T=0$ tricritical points (where the
N-AMSF transition changes order near $\delta \simeq 3$~meV, and where
the MSF-AMSF transition changes order near $\delta \simeq
7$~meV).  At such points the finite temperature transition switches
  from being very weakly first order (controlled by the polariton
  mass) to being strongly first order, due to the resonance. As the
  temperature increases, while the lower ``tricritical point'' does
  not change substantially, the triple point approaches the upper
  ``tricritical point'', so that, for $T \simeq 2.32$~K, the
  ``tricritical point'' and triple point have merged, as seen in the
  bottom panel of the S.~Fig.~\ref{fig:gaas-vs-detun}.

The cuts at fixed $\delta$ in S.~Fig.~\ref{fig:gaas-vs-T} show the
same
evolution that we have just described at fixed $T$. The size of jump
in density at the AMSF-MSF transition is determined by whether the
detuning is above or below the $T=0$ tricritical point. At
$\delta=6.5$~meV (which, at $T=0$, is in between the critical
end-point and the upper tricritical point), there is a first order
transition between MSF-AMSF even at zero temperature. The panel at
$\delta=8.3$~meV is very close to (and slightly above) the $T=0$ upper
tricritical point, so the MSF-AMSF transition is second order at zero
temperature and evolves to strongly first order at higher
temperatures. Finally, for $\delta=10$~meV, the zero temperature
transition is second order and becomes (weakly) first order at
non-zero temperature.

\begin{figure}
\centering
\includegraphics[width=3.2in]{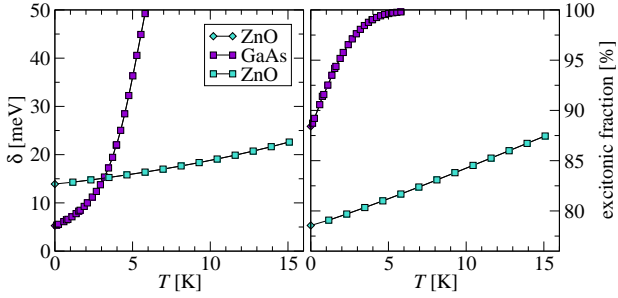}
\caption{Location of the triple points as a function of temperature
  for GaAs and ZnO parameters in the low-momentum quadratic
    approximation for the LP dispersion --- note that at $T=0$ the
  triple points (squares) evolves into a critical end-point (diamond).
  Left panel: required detuning.  Right panel: equivalent excitonic
  fraction.}
\label{fig:triple-point}
\end{figure}
%
\subsection{Evolution of triple point}
\label{sec:evol-triple-point}
The finite temperature phase diagrams presented above indicate that
the MSF region shrinks on either reducing the detuning $\delta$ or on
increasing the temperature $T$. This behavior can be quantified by
evaluating the location of the triple point at which N, MSF and AMSF
phases all meet. In order for an MSF phase to be possible, one must
have a lower temperature, or a larger $\delta$ than that of the triple
point.  Supplemental Figure~\ref{fig:triple-point} shows the evolution
of the detuning of the triple point vs temperature --- note that at
$T=0$ the triple points (squares) evolves into a critical end-point
(diamond).  We show results both for GaAs parameters as discussed
above, as well as for ZnO parameters discussed below.  One may use the
Hopfield coefficient $c_{0}^2$ to translate a detuning into an
excitonic fraction; this is shown in the right panel of
S.~Fig.~\ref{fig:triple-point}.  It is clear from this figure that
while the MSF phase can be reached for the commonly used microcavities
made of GaAs, it requires relatively low temperatures, and high
excitonic fractions. In ZnO, however, the conditions are significantly
more favorable. It should however be noted that comparing Fig.~3 of
the manuscript to the $\delta=10$~meV panel in
S.~Fig~\ref{fig:gaas-vs-T} shows that the quadratic approximation may
significantly underestimate the critical temperature. This suggests
experimental parameters may be even more favorable than
S.~Fig.~\ref{fig:triple-point} indicates.

\begin{figure}
  \centering
  \includegraphics[width=3.2in]{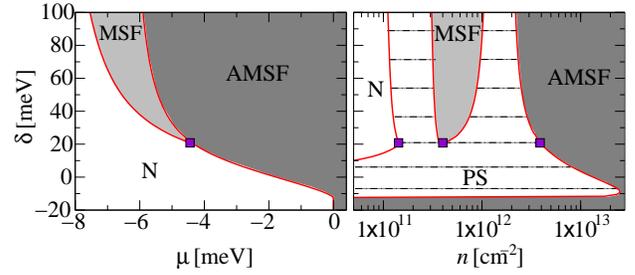}  
  \caption{Finite temperature phase diagram for ZnO parameters in
      the low-momentum quadratic approximation for the LP dispersion,
    as discussed below, for a temperature $T=11.6$~K. Note that a
    logarithmic scale has been used for density in the right hand
    panel.}
  \label{fig:zno-vs-delta}
\end{figure}
%
\subsection{ZnO parameters}
\label{sec:zno-parameters}
For ZnO, while the photon and exciton masses are similar to those of
GaAs, $m_C = 10^{-4} m_X$ and $m_X = 0.2 m_e$ (for the exciton mass
see the Supplementary Information of Ref.~\cite{Li2013a}), the most
important change is the significant increase of biexciton binding
energy to $|E_b|=18$~meV~\cite{Makino2005} (also the exciton binding
energy increases from $10$~meV for GaAs to around $85$~meV, depending
on the well width). The Rabi-Splitting for ZnO also is typically
larger than for GaAs, reaching values up to
$200-300$~meV~\cite{Li2013a}.
However, for the observation of the MSF phase, it is in fact
advantageous to consider a smaller Rabi splitting than the largest
values that have been achieved~\cite{Li2013a,Trichet2013}, and we fix
$\Omega_R=20$~meV. Reducing the Rabi splitting is easy to achieve, by
using a longer microcavity, with a higher mode volume. We hence choose
a reasonable value where the MSF phase occurs at easily attainable
detunings and temperatures.  Supplemental
Figure~\ref{fig:zno-vs-delta} shows a characteristic phase diagram of
ZnO at a relatively high temperature of $T = 11.6$~K, showing that the
MSF phase can in this case be seen for high temperatures and detunings
comparable to the Rabi splitting.


%

\end{document}